\documentstyle[preprint,aps,epsf]{revtex}

\def\ssize{\scriptsize}  

\tightenlines
\catcode`@=11
\def\references{%
\ifpreprintsty
\bigskip\bigskip
\hbox to\hsize{\hss\large \refname\hss}%
\else
\vskip24pt
\hrule width\hsize\relax
\vskip 1.6cm
\fi
\list{\@biblabel{\arabic{enumiv}}}%
{\labelwidth\WidestRefLabelThusFar  \labelsep4pt %
\leftmargin\labelwidth %
\advance\leftmargin\labelsep %
\ifdim\baselinestretch pt>1 pt %
\parsep  4pt\relax %
\else %
\parsep  0pt\relax %
\fi
\itemsep\parsep %
\usecounter{enumiv}%
\let\p@enumiv\@empty
\def\theenumiv{\arabic{enumiv}}%
}%
\let\newblock\relax %
\sloppy\clubpenalty4000\widowpenalty4000
\sfcode`\.=1000\relax
\ifpreprintsty\else\small\fi
}
\catcode`@=12

\begin{document}

\font\fortssbx=cmssbx10 scaled \magstep2
\hbox to \hsize{
\hbox{\fortssbx University of Wisconsin - Madison}
\hfill\vtop{\hbox{\bf MADPH-96-1019}
                \hbox{\bf IFUSP-P.1281}
                \hbox{\bf hep-ph/9710443}} }

\vspace{.5in}

\begin{center}
{\Large\bf  The Associated Production of Weak Bosons and Jets by Multiple
Parton Interactions}\\[4mm]
O. \'Eboli$^a$, F. Halzen$^b$, and J. K. Mizukoshi$^a$
\end{center}

\begin{center}
\it
$^a$ Instituto de F\'{\i}sica, Universidade de S\~ao Paulo,\\
 C.P. 66.318, CEP 05315-970 S\~ao Paulo, Brazil. \\[1.5mm]
$^b$ Physics Department, University of Wisconsin, Madison, WI 53706.
\end{center}

\thispagestyle{empty}  

\vspace{.2in}

\begin{abstract}
The sources of $W + n$-jet events in hadron collisions are higher-order QCD
processes, but also multiple-parton interactions.  A
subprocess producing a $W + k$-jet final state, followed by one
producing $l$ jets in the same nucleon-nucleon interaction, will
result in a $W + n$-jet event if $k + l = n$. In the simplest case a
$W + 2$-jet event can be produced by a quark-antiquark annihilation
into $W$ and a 2-jet event occurring in the same proton-antiproton
interaction. We compute that this happens at the 10\% level of the
higher-order QCD processes for the type of cuts made by the Tevatron
experiments. For jet $p_T$ values of order $5{\sim}10$~GeV, multiple-parton
interactions dominate higher-order QCD-processes. The emergence of
this new source of $W + n$-jet events towards lower $p_T $ simulates
the running of $\alpha_s$; it is imperative to remove these
processes from the event sample in order to extract information on
the strong coupling constant. Also, BFKL studies of low-$p_T$ jet cross
sections are held hostage to a detailed understanding of the multiple-parton
interactions. We perform the calculations required to achieve these goals. A
detailed experimental analysis of the data
may, for the first time, determine the effective areas occupied by
quarks and gluons in the nucleon. These are not necessarily
identical. We also compute the multiple-parton contribution to $Z +
n$-jet events.
\end{abstract}

\newpage


\section{Introduction}

The very clean samples of $W, Z + n$-jet events produced in
proton-antiproton interactions at the Tevatron \cite{cdf:wz} have
become a laboratory to study QCD \cite{cdf:qcd}. In principle, these
events can be used to determine the strong coupling $\alpha_s$. In
this paper we point out that the same events can be used to study
multiple-parton interactions. It would, in fact, be treacherous to
determine $\alpha_s$ without removing their contribution to the data
sample. We will show that their contribution strongly increases
relative to the higher-order QCD processes with decreasing values of
$p_T$. This behavior obviously mimics the running of the strong
coupling. Furthermore, studies of BFKL physics with jets cover the low-$p_T$
region where multiple-parton processes eventually dominate and
will have to await a detailed understanding of these processes.

\begin{figure}
\centering
\leavevmode
\epsfxsize=5in\epsffile{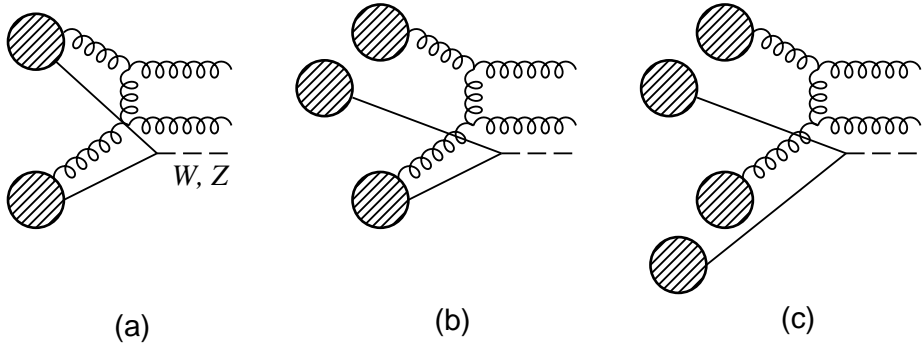}

\caption{$W,Z + 2$-jet multiple-parton scattering.}
\label{fig:mult}
\end{figure}

We have computed the contributions of multiple interactions and higher-order
QCD processes to the processes $W, Z + 2$-jets; see
Fig.~\ref{fig:mult}. In the simplest case of a $W + 2$-jet event the
final state can be produced by a quark-antiquark annihilation into $W$
and a 2-jet event, both occurring in a single beam crossing
\cite{ant}. The two-parton subprocess may involve the same, or
different beam particles.  In most cases the latter process can be
identified by the detector and removed from the data set
\cite{cdf:dp}. Our conclusions are that: i)~higher-order QCD processes
and multiple-parton interactions are similar in magnitude for $p_T$
values of the jets of order $5{\sim}10$~GeV and ii)~even for jets with
a minimum $p_T$ of 20~GeV the multi-parton contribution is at the
$10{\sim}20\%$ level. Notice that a contribution of this magnitude is
sufficiently important to interfere with a determination of
$\alpha_s$. The obvious way to remove it is to study the event sample
as a function of the minimum $p_T$ of the jets.  We have also
calculated the multiple interaction and higher-order QCD contributions
to $W + 3$-jet event and comment on their relative importance when the
number of jets produced in association with the $W$ increases.
Finally, we repeated the calculations for $Z + 2,3$-jet events.

Multi-parton scattering is of theoretical interest because it probes
the partonic structure of hadrons in novel ways. The joint probability
that a quark and antiquark annihilate into a $W$, and that another
pair of partons produces a 2-jet final state in a single
proton-antiproton interaction is
\begin{equation}
\sigma_{W\,\mbox{\ssize 2-jet}} = \sigma_W\left(\sigma_{\mbox{\ssize
2-jet}}\over \pi R^2\right)\,.
\label{eq:joint}
\end{equation}
The result is obvious except for the factor ${1 \over \pi R^2}$. It
takes into account the fact that, once the first parton process has
taken place, the area of the beam has been reduced to the area of the
proton in which the first interaction occurred. In principle, there
could exist correlations in the double-parton-scattering
structure functions that would invalidate this expression. For
instance, we are neglecting the correlation due to
longitudinal momentum conservation, which should introduce factors of
the form $(1-x_1-x_2)$, with $x_i$ being the fraction of momentum
carried by the parton $i$.  In the small $x$ regions, covered in our
calculations, Eq.~(\ref{eq:joint}) is valid.  The effective area $\pi R^2$
should be of order of the inelastic cross section, or about 50~mb.  Earlier
studies of final states with multiple jets and
multiple Drell-Yan pairs~\cite{early}, although admittedly plagued by
difficult systematics or low statistics, have agreed on the much
smaller value of roughly 15~mb. This may indicate that hard quarks and
gluons only occupy a small area of the high-energy
proton. Anticipating the higher quality of the Tevatron data, it may
ultimately be possibly to separately measure the areas occupied by
quarks and gluons. Several authors have speculated that gluons may be
clustered in the vicinity of valence quarks~\cite{hwa}.


\section[]{ \boldmath $W + 2$-jets event rates}

The multiple-interaction cross section for the production of a final state
$A$ in association with the state $B$, is given by
\begin{equation}
  \sigma_{A-B}
 = \sigma_A \sigma_B \left[
    {1\over\pi R^2}
    + 2 \sqrt{ {1\over\pi R^2} \, {(N-1)\over\sigma_{\rm inel}} } +
            {(N-1)\over\sigma_{\rm inel}}
  \right] \;\; ,
\label{gen}
\end{equation}
for $A\ne B$. Here $N$ stands for the average number of interactions
per beam crossing. Summation over $A$ and $B$ states leading to the
same $A-B$ final state is understood. The first term corresponds to
double-parton interactions involving the same beam particles, like in
Eq.~(\ref{eq:joint}), while the last term describes simultaneous
interactions between different beam particles. As for the second term in
Eq.~(1), it corresponds to the interaction of two partons in a proton in one
bunch with two different protons in the other bunch. It can be understood as
the interference term for the first two reaction mechanisms. In our calculation
we use $R =0.7$~fm, which corresponds to an effective cross section $\pi
R^2= 15.4$~mb, $\sigma_{\rm inel} = 51.7$~mb, and $N = 1.8$.  This value
of $R$ is also consistent with the the one derived from the analysis
of the $\gamma + 3 $-jet production by CDF~\cite{cdf:dp}.

In order to obtain the multiple-parton and QCD cross sections for the
production of a $W$ and two jets, we used the package
MADGRAPH~\cite{tim} to evaluate the relevant tree-level matrix
elements. These agree with previous evaluations. We imposed a
minimum-$p_T$ cut on the transverse momenta of the jets and required that the
jets (partons) be in the rapidity region $| \eta_j| < 3$, and
separated by $\Delta R = \sqrt{\Delta \eta^2 + \Delta \phi^2} \ge
0.9$. We used the CTEQ3M structure functions \cite{cteq} and evaluated
the QCD scales at $\hat{s}/2$, where $\hat{s}$ is the subprocess
center-of-mass energy.

\begin{figure}
\centering
\leavevmode
\epsfxsize=5in\epsffile{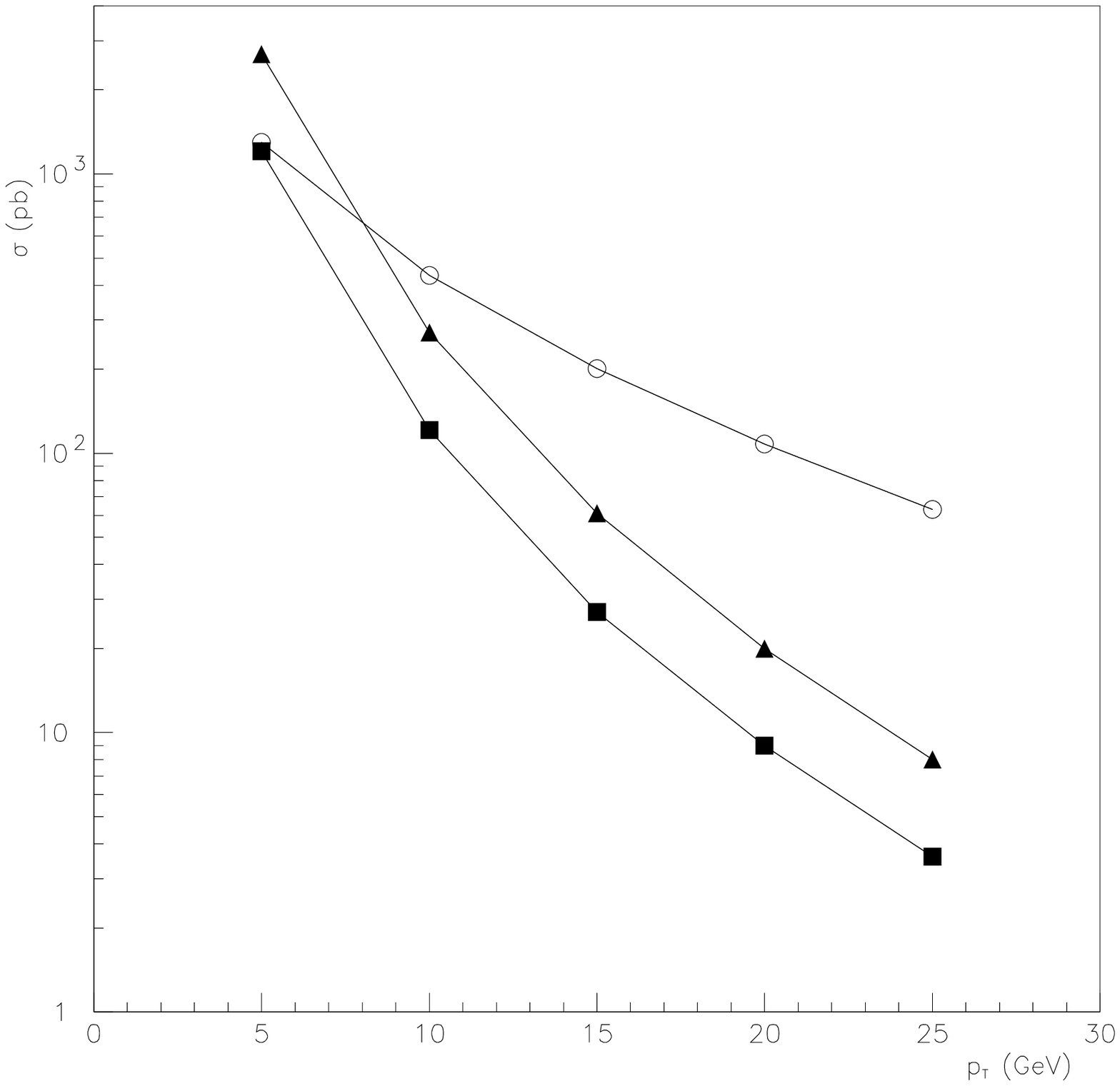}

\caption{ Total cross section for the production of a $W$ accompanied
  by 2 jets, as a function of the minimum-$p_T$ cut, for:  QCD processes
  (open circles), double-parton interactions (triangles), multiple
  interactions (squares).}
\label{fig:w2s}
\end{figure}

\begin{figure}
\centering\leavevmode
\epsfxsize=3.5in\epsffile{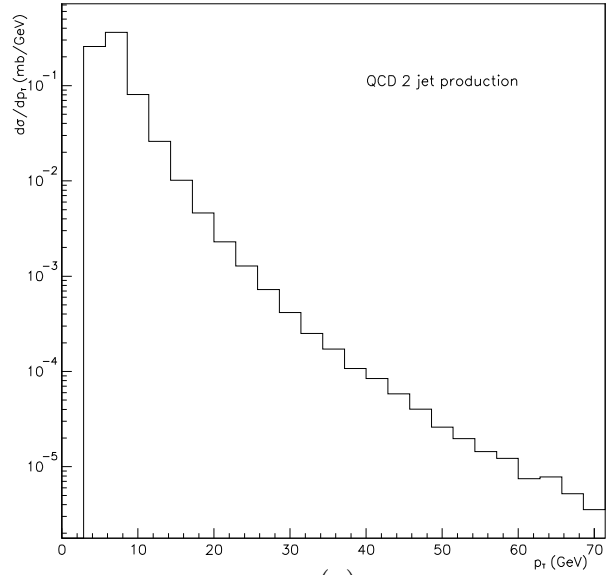}
\begin{center}\vskip-.4in
(a)
\end{center}
\leavevmode
\epsfxsize=3.5in\epsffile{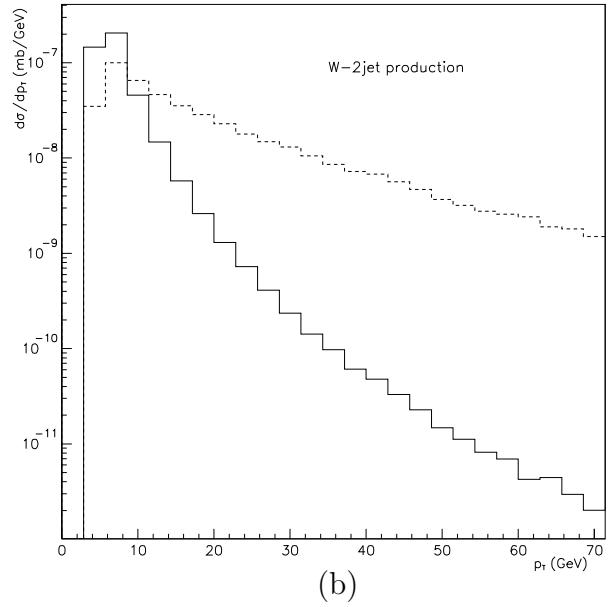}
\begin{center}\vskip-.4in
(b)
\end{center}
\caption{ $p_T$ distribution of the jets in mb/GeV, for: (a) QCD 2-jets
  production, (b) $W+2$-jets production.}
\label{fig:ptw2j}
\end{figure}

Our results for the $W+2$ jets production are exhibited in
Fig.~\ref{fig:w2s} as a function of the minimum-$p_T$ cut. We present
in Fig.~\ref{fig:ptw2j} the $p_T$ spectrum of the jets produced in
association with the $W$ for both multiple-interaction and higher-order QCD
processes. As can be seen from this figure, the jets
produced in multi-parton scattering are much softer than those
produced by higher-order processes. This is why multiple-particle
interactions become more important with decreasing $p_T$.

\begin{figure}
\centering
\leavevmode
\epsfxsize=5in\epsffile{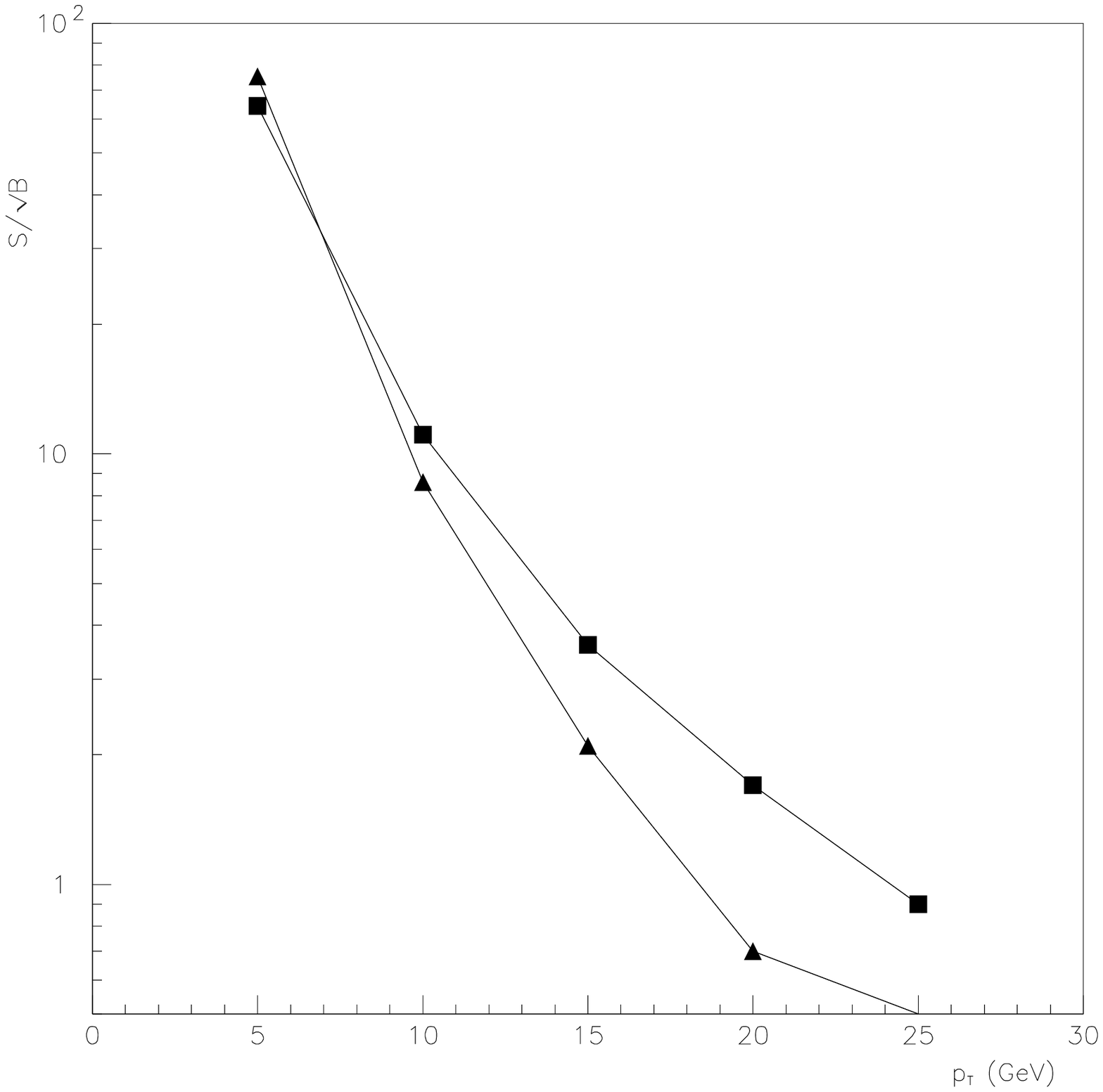}

\caption{Statistical significance (Signal/$\protect\sqrt{\rm background}$) of
the double-parton interactions for the production of a $W+2$-jets (squares) and
$W+3$-jets (triangles) as a function of the minimum
transverse momenta of the jets.}
\label{fig:st2j}
\end{figure}

In Fig.~\ref{fig:st2j} we plot the statistical significance (signal
over square root of background) for multi-parton interactions assuming
an integrated luminosity of 100~pb$^{-1}$, Br$(W \rightarrow e \nu_e)
= 10.8$\%, and an efficiency of 34\% to detect the $W$ \cite{cdf:wz}.
The double-parton interactions can be observed in the $W+2$-jet final
state with a significance in excess of 5$\sigma$ provided the minimum
$p_T$ of the jets is smaller than 10--12 GeV. Clearly, the signal for
double-parton scattering can be further enhanced, e.g.\ by a cut
requiring that the jets are back-to-back in the transverse plane in a
$W+2$ jet event.  The poorly known effective area $1 \over {\pi R^2}$
dominates the uncertainty of the calculations. The results shown
assumed a common value of 15.4~mb for quarks and gluons, corresponding
to $R = 0.7$~Fermi. It is clear from these results that the Tevatron
has the opportunity to determine this quantity with much improved
precision.


\section[]{\boldmath $W + 3$-jets event rates}

\begin{figure}
\centering
\leavevmode
\epsfxsize=5in\epsffile{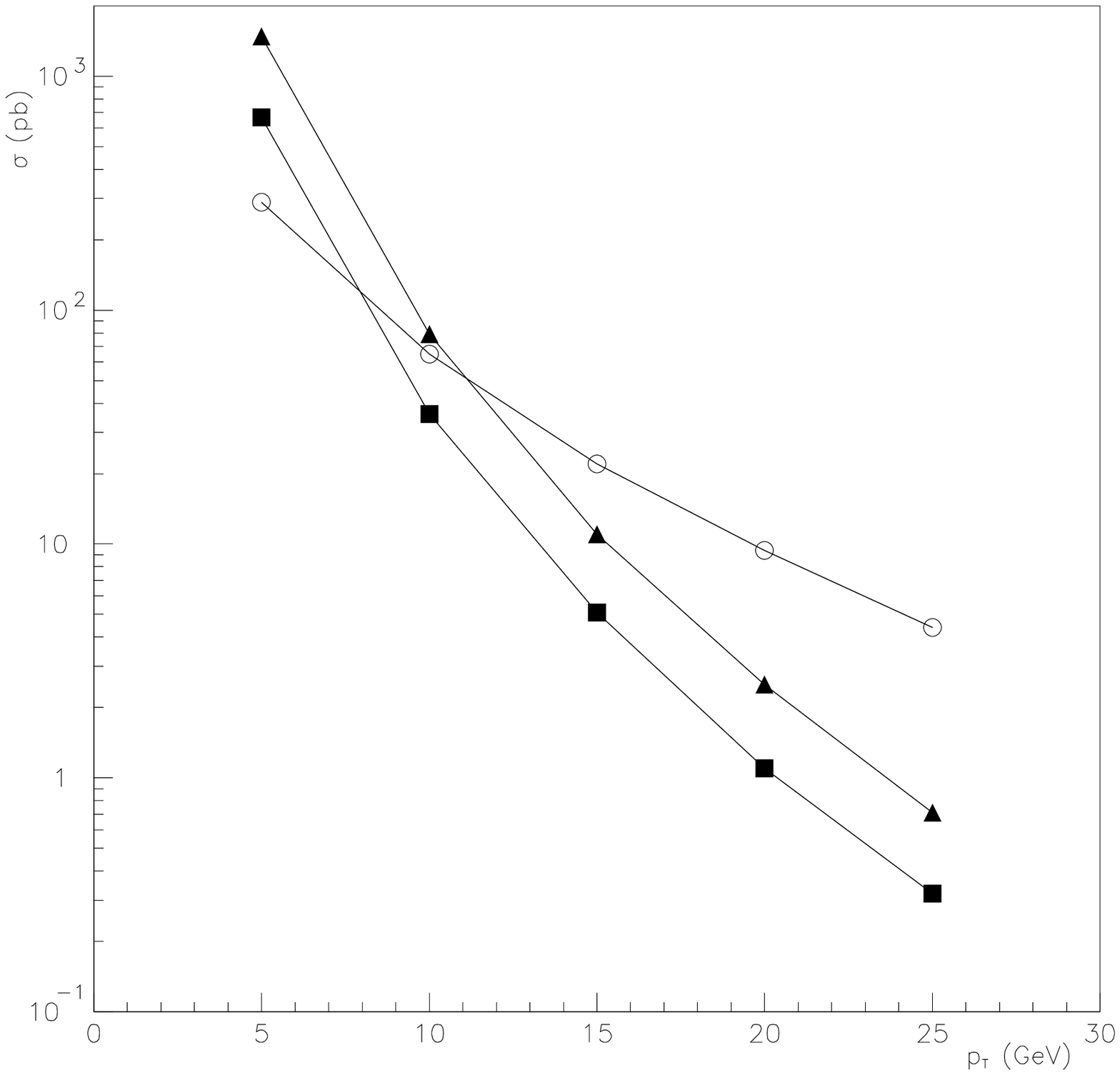}

\caption{ Total cross section for the production of a $W$ accompanied
by 3 jets, as a function of the minimum-$p_T$ cut, for: higher-order
QCD processes (open circles), double-parton interactions
(triangles), multiple interactions (squares).}
\label{fig:w3s}
\end{figure}

\begin{figure}
\centering\leavevmode
\epsfxsize=3.5in\epsffile{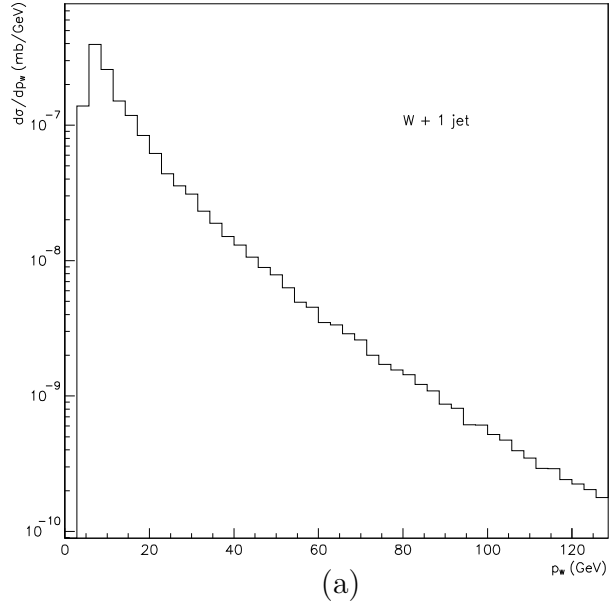}
\begin{center}\vskip-.4in
(a)
\end{center}
\leavevmode
\epsfxsize=3.5in\epsffile{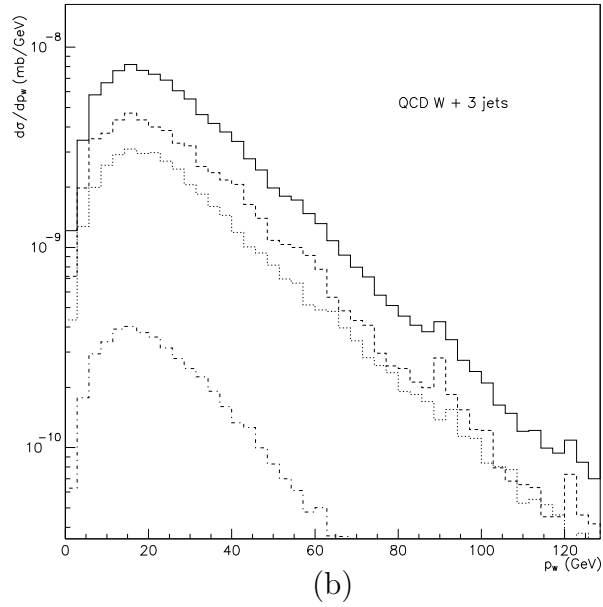}
\begin{center}\vskip-.4in
(b)
\end{center}
\caption{ $p_T$ distribution of the $W$ in mb/GeV, for: (a) $W+1$-jet
production;
(b)~QCD $W+3$-jets production with the solid line standing for the
total while the dashed, dotted, and dot-dashed lines stand for the
contribution from the quark-gluon, quark-quark, and gluon-gluon
subprocesses respectively .}
\label{fig:ptw3j}
\end{figure}

Double-parton interactions can also be studied in the production of $W+3$-jets.
The cross section for this process is given by Eq.~(\ref{gen}) with
\[
\sigma_A \sigma_B =
        \sigma_{W\,\rm jet}\sigma_{\mbox{\ssize 2-jet}} +
\sigma_W\sigma_{\mbox{\ssize 3-jet}} \; .
\]
The cross sections for this final state via higher-order QCD,
double-parton interactions and multi-beam interactions are shown in
Fig.~\ref{fig:w3s} as a function of the minimum $p_T$ of the jets.
As we can see from Fig.~\ref{fig:ptw3j}, the transverse momenta of
the $W$ produced via higher-order QCD are larger than those generated
in the double-parton interaction. This suggests that a cut like $p_T^W
\alt 20$ GeV should increase the signal-to-background ratio.

\begin{figure}
\centering
\leavevmode
\epsfxsize=5in\epsffile{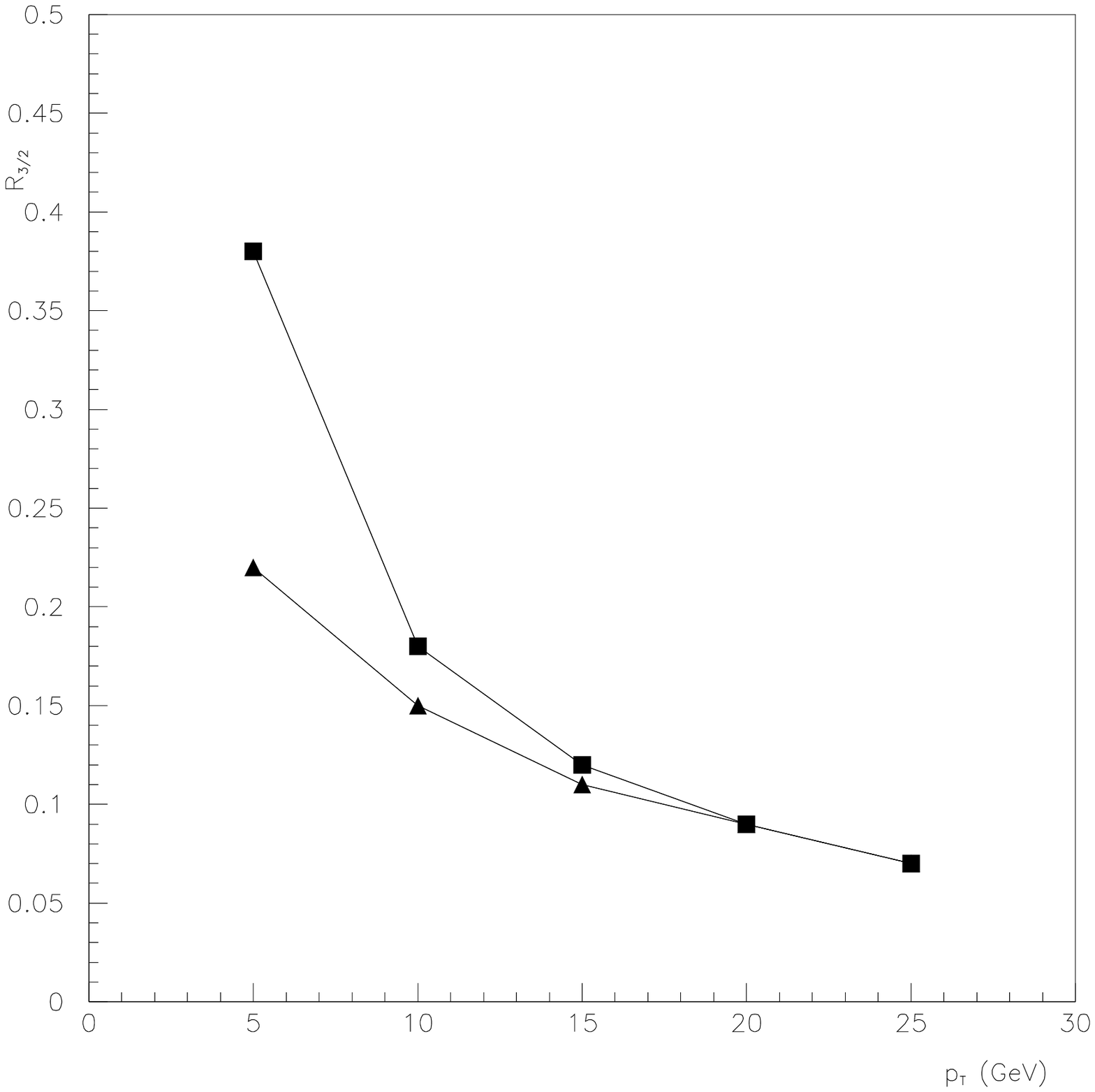}

\caption{ $R_{3/2}$ in $W+n$-jet production as a function of the jet
  minimum transverse momentum. The squares stand for the sum of the
  QCD higher-order and double-parton contributions, while the
  triangles represent only the higher-order QCD contributions.}
\label{fig:r32}
\end{figure}

Figure \ref{fig:st2j} also shows the statistical significance for the
double-parton production of $W+3$-jets as a function of the minimum
transverse momenta of the jets, indicating that the signal should be
visible for $p^{\rm cut}_T \alt 12$ GeV. This curve was obtained using
the same luminosity and efficiencies assumed for the $W+2$-jet
signal. No further cuts were introduced to enhance the signal.  Notice
that double-parton scattering is observable in both $W+2$-jet and
$W+3$-jet channels for approximately the same range of transverse
momenta.  Therefore, the observation of a signal in one channel can be
confirmed in the other. Moreover, it is possible to observe the
presence of double-parton scattering by analyzing the ratio $R_{3/2} =
\sigma_{ W+\mbox{\ssize 3-jets}}/\sigma_{ W+\mbox{\ssize 2-jets}}$.  As shown
in Fig.~\ref{fig:r32}, double-parton interactions enhance this ratio by a
factor 1.5--2 for low-$p_T$ cuts.


%
%

\section[]{\boldmath $Z+\lowercase{n}$-jets}

Multiple-parton scattering can also be studied through the production
of a $Z$ accompanied by 2 or 3 jets. In order to study these processes
we must evaluate
\begin{eqnarray}
\left. \sigma_A \sigma_B  \right |_{Z\, \mbox{\ssize 2-jet}} &=&
        \sigma_Z\sigma_{\mbox{\ssize 2-jet}}  \;\; ,
\\
\left . \sigma_A \sigma_B \right |_{Z\, \mbox{\ssize 3-jet}}&=&
        \sigma_{Z\,\rm jet}\sigma_{\mbox{\ssize 2-jet}} +
\sigma_Z\sigma_{\mbox{\ssize 3-jet}} \;\; .
\end{eqnarray}
We exhibit in Table~\ref{tab:z2j} (\ref{tab:z3j}) the cross sections
for the production of $Z+2$-jets ($Z+3$-jets) from higher-order QCD
processes, double-parton scattering and multiple-beam interactions.
These results were obtained imposing a minimum-$p_T$ cut on the
transverse momenta of the jets. As before, we required that the jets
(partons) populate the rapidity region $| \eta_j| < 3$, separated by
$\Delta R \ge 0.9$.

\begin{figure}
\centering\leavevmode
\epsfxsize=5in\epsffile{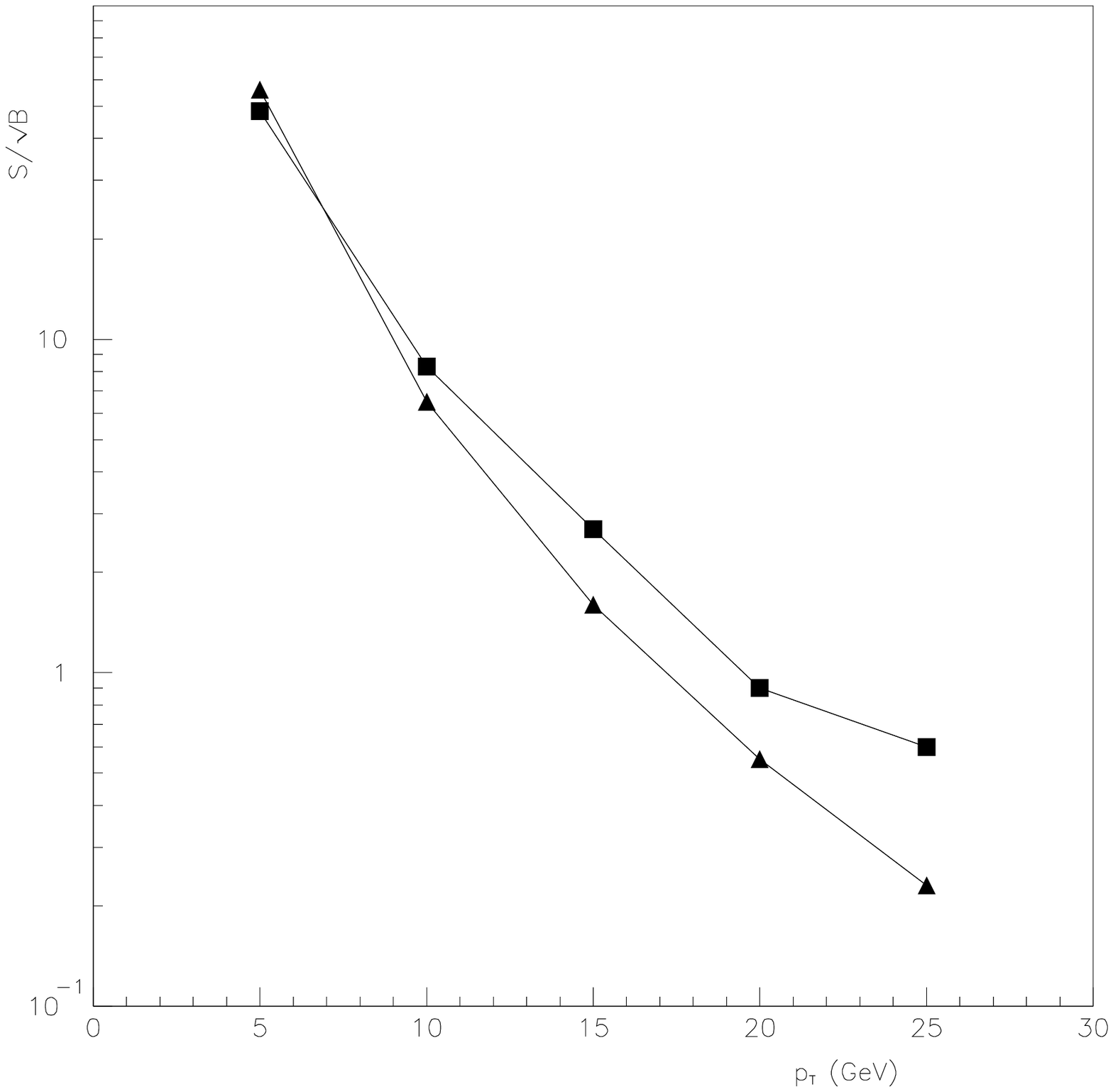}

\caption{Statistical significance (Signal/$\protect\sqrt{\rm background}$)
  of the double-parton interactions for the production of a $Z+2$-jets
  (squares) and $Z+3$-jets (triangles) as a function of the minimum
  transverse momenta of the jets.}
\label{fig:stz2j}
\end{figure}

\begin{figure}
\centering
\leavevmode
\epsfxsize=5in\epsffile{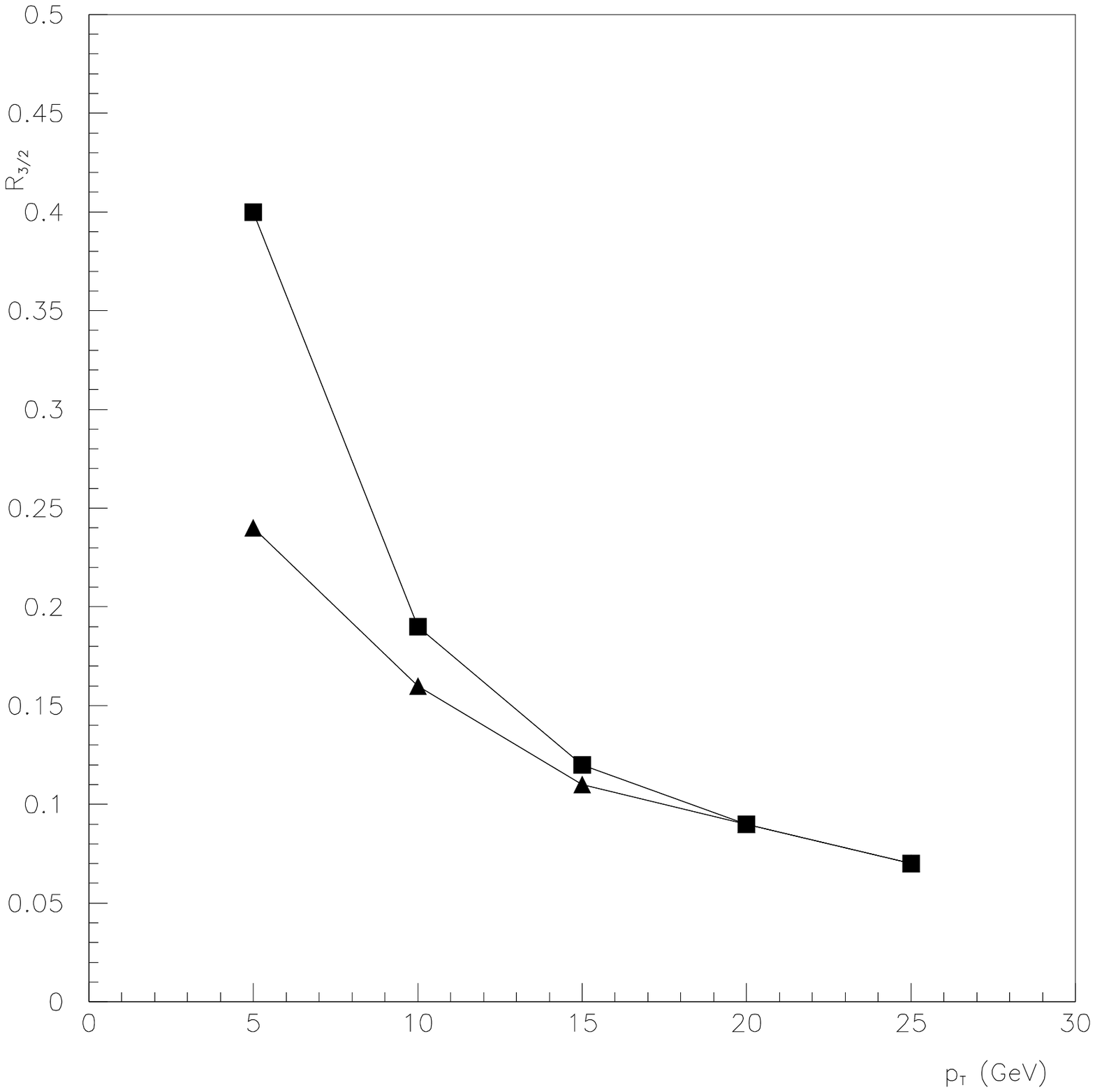}

\caption{ $R_{3/2}$ in $Z+n$-jet production as a function of
  the jet minimum transverse momentum. The squares stand for the sum
  of the QCD higher-order and double-parton contributions, while the
  triangles represent only the higher-order QCD contributions.}
\label{fig:r32z}
\end{figure}

The statistical significance of the multi-parton contribution to the
$Z+2$-jet production is shown in Fig.~\ref{fig:stz2j} for $Z$ decaying
into $e^+e^-$ or $\mu^+\mu^-$ pairs, and an integrated luminosity of
100 pb$^{-1}$. The $Z$ acceptance was taken to be 41\%
\cite{cdf:wz}. We conclude that the multi-parton scattering can be
observed at a 3$\sigma$ level for $p^{\rm cut}_T \le 14$~GeV. The signal
should be very clean for low values of the $p_T$ cut, even without
further cuts requiring the $Z$ and jets to have pair-wise balancing
transverse momenta. The signal for multi-parton scattering in the
production of a $Z+ 3$-jets is also visible for a $p_T$ cut smaller
than 13~GeV, as shown in Fig.~\ref{fig:stz2j}. Analogously to the
$W+n$-jets case, the signal for multi-parton interactions in $Z+n$-jets also
appears in the ratio $R_{3/2} = \sigma_{ Z+\mbox{\ssize 3-jets}}/\sigma_{
Z+\mbox{\ssize 2-jets}}$, as shown in Fig.~\ref{fig:r32z}.

\section*{Acknowledgements}
This research was supported in part by the U.S.~Department of Energy under
Grant No.~DE-FG02-95ER40896, in part by the University of Wisconsin Research
Committee with funds granted by the Wisconsin Alumni Research Foundation, and
in part  by Conselho Nacional de Desenvolvimento
Cient\'{\i}fico e Tecnol\'ogico (CNPq), and by Funda\c{c}\~ao de
Amparo \`a Pesquisa do Estado de S\~ao Paulo (FAPESP).



\newpage

\begin{table}
\caption{$Z  + 2$-jet cross sections at the Fermilab Tevatron, with the
cuts and parameters discussed in the text.}
\label{tab:z2j}
\begin{center}
\begin{tabular}{|c|c|c|c|}
\hline
$p_T$ cut (GeV) &
$\sigma^{\rm QCD}_{Z\, \mbox{\ssize 2-jet}}$ (pb)&
$\sigma_{Z\, \mbox{\ssize 2-jet}}^{\rm mult.\ int.}$ (pb) &
$\sigma_{Z\, \mbox{\ssize 2-jet}}^{\rm mult.\ part.}$ (pb) \\
\hline
\hline
5  & 800. & 1578. & 710. \\
\hline
10 & 269. & 159.  & 71.  \\
\hline
15 & 126. & 36.   & 16.  \\
\hline
20 & 67.  & 12.   & 5.3  \\
\hline
25 & 40.  & 4.7   & 2.1  \\
\hline
\end{tabular}
\end{center}
\end{table}

\vskip1in

\begin{table}
\caption{$Z + 3$-jet cross sections at the Fermilab Tevatron, with the cuts and
parameters discussed in the text.}
\label{tab:z3j}
\begin{center}
\begin{tabular}{|c|c|c|c|c|}
\hline
$p_T$ cut (GeV) &
$\sigma^{\rm QCD}_{Z\,\rm jet} $ (pb) &
$\sigma_{Z\, \mbox{\ssize 3-jet}}$ (pb) &
$\sigma_{Z\, \mbox{\ssize 3-jet}}^{\rm mult.\ int.}$ (pb) &
$\sigma_{Z\, \mbox{\ssize 3-jet}}^{\rm mult.\ part.}$ (pb) \\
\hline
\hline
5  & 2582. & 195.  & 906. &  408.   \\
\hline
10 & 1386. & 42.  & 49.  &  22.   \\
\hline
15 & 868.  & 14.5  & 7.2  &  3.2   \\
\hline
20 & 586.  & 6.2  & 1.6  &  0.72   \\
\hline
25 & 412.  & 3.0  & 0.46 &  0.21  \\
\hline
\end{tabular}
\end{center}
\end{table}


\clearpage

\end{document}